\documentclass{elsart}
\usepackage{graphicx}
\usepackage{amssymb}

\begin{document}

\begin{frontmatter}

\title{Beam Energy and Centrality Dependence of Two-Pion Bose-Einstein 
Correlations at SPS Energies}

\author[rez]{D.~Adamov\'a},
\author[gsi]{G.~Agakichiev}, 
\author[hei]{H.~Appelsh\"auser}, 
\author[dub]{V.~Belaga}, 
\author[gsi]{P.~Braun-Munzinger}, 
\author[hei]{R.~Campagnolo},
\author[gsi]{A.~Castillo},
\author[wei]{A.~Cherlin}, 
\author[hei]{S.~Damjanovi\'c}, 
\author[hei]{T.~Dietel}, 
\author[hei]{L.~Dietrich}, 
\author[sun]{A.~Drees}, 
\author[hei]{S.\,I.~Esumi}, 
\author[hei]{K.~Filimonov}, 
\author[dub]{K.~Fomenko},
\author[wei]{Z.~Fraenkel}, 
\author[gsi]{C.~Garabatos}, 
\author[hei]{P.~Gl\"assel}, 
\author[gsi]{G.~Hering}, 
\author[gsi]{J.~Holeczek}, 
\author[rez]{V.~Kushpil}, 
\author[cer]{B.~Lenkeit}, 
\author[hei]{W.~Ludolphs},
\author[gsi]{A.~Maas}, 
\author[gsi]{A.~Mar\'{\i}n}, 
\author[hei]{J.~Milo\v{s}evi\'c},
\author[wei]{A.~Milov}, 
\author[gsi]{D.~Mi\'skowiec}, 
\author[cer]{L.Musa},
\author[dub]{Yu.~Panebrattsev}, 
\author[dub]{O.~Petchenova}, 
\author[hei]{V.~Petr\'a\v{c}ek}, 
\author[cer]{A.~Pfeiffer}, 
\author[mpi]{J.~Rak}, 
\author[wei]{I.~Ravinovich}, 
\author[bnl]{P.~Rehak}, 
\author[hei]{M.~Richter}, 
\author[gsi]{H.~Sako}, 
\author[hei]{W.~Schmitz}, 
\author[cer]{J.~Schukraft}, 
\author[gsi]{S.~Sedykh}, 
\author[hei]{W.~Seipp}, 
\author[gsi]{A.~Sharma}, 
\author[dub]{S.~Shimansky}, 
\author[hei]{J.~Sl\'{\i}vov\'a},
\author[hei]{H.\,J.~Specht}, 
\author[hei]{J.~Stachel}, 
\author[rez]{M.~\v{S}umbera}, 
\author[hei]{H.~Tilsner}, 
\author[wei]{I.~Tserruya}, 
\author[gsi]{J.\,P.~Wessels}, 
\author[hei]{T.~Wienold}, 
\author[hei]{B.~Windelband}, 
\author[mpi]{J.\,P.~Wurm}, 
\author[wei]{W.~Xie}, 
\author[hei]{S.~Yurevich}, 
\author[dub]{V.~Yurevich}

\begin{center}
(CERES Collaboration)
\end{center}

\address[rez]{NPI ASCR, \v{R}e\v{z}, Czech Republic}
\address[gsi]{GSI Darmstadt, Germany}
\address[hei]{Heidelberg University, Germany}
\address[dub]{JINR Dubna, Russia}
\address[wei]{Weizmann Institute, Rehovot, Israel}
\address[sun]{SUNY at Stony Brook, U.S.A.}
\address[cer]{CERN, Geneva, Switzerland}
\address[bnl]{BNL, Upton, U.S.A.}
\address[mpi]{MPI, Heidelberg, Germany}

\begin{abstract}
Results are presented of a two-pion interferometry (HBT) analysis
in Pb+Au collisions at 40, 80, and 158 AGeV.
A detailed study of the Bertsch-Pratt HBT radius parameters  
has been performed
as function of the mean pair transverse momentum $k_t$ and in bins
of the centrality of the collision. 
From these results we extract 
model dependent information about the space-time evolution 
of the reaction. An investigation of the effective
volume of the pion emitting system provides an important
tool to study the properties of thermal pion freeze-out. 
\end{abstract}

\end{frontmatter}

\section{INTRODUCTION}
Lattice QCD calculations predict a transition from confined hadronic matter
to a state of deconfined quarks and gluons at a critical energy density 
$\epsilon_{c}$ around 0.6~GeV/fm$^{3}$~\cite{lattice}. 
Such energy densities are believed
to be reached at CERN SPS energies and indeed evidence for a new state of matter
created in Pb-induced reactions at the SPS has been announced recently~\cite{success}.   
It is expected that the reaction dynamics could differ considerably if energy
densities close to or above $\epsilon_{c}$ are reached, and that the
existence of a phase transition could manifest itself in the space-time evolution
of the system~\cite{hushu,ber,pra,bergong,berbrown,rischke,rigyu}.
In particular, the beam energy dependence of quantities characterizing
the space-time evolution may give hints to the onset of deconfinement
in heavy ion collisions. In addition, the beam energy range available at the SPS  
provides an important link between the nucleon dominated domain at the AGS 
and the pionic regime at collider energies,  
leading to a better understanding of the mechanisms relevant
for thermal freeze-out.

While single particle momentum distributions are also determined by
the space-time evolution, the lifetime and the spatial extent of the
system as well as the existence of collective velocity fields at the time
of thermal freeze-out can be disentangled by the study of Bose-Einstein (BE) momentum  
correlations of identical pions via Hanbury-Brown and Twiss interferometry (HBT). 
The width of
the correlation peak at vanishing relative momenta reflects the
so-called length of homogeneity of the pion emitting source. 
Only in static sources can the length of homogeneity, in the following
also called `source radius', be interpreted
as the true geometrical size of the system.
In a dynamic system, the occurence of space-momentum correlations of the 
emitted particles due to collective expansion generally leads to
a reduction of the observed source radii. The degree of reduction depends on
the gradients of the collective expansion velocity and the thermal velocity of the
pions 
at thermal freeze-out. 
A differential analysis of the HBT correlations in bins
of the pair transverse momentum 
thus provides valuable information not only on the spatial extent but also on
the properties of the collective
expansion of the system~\cite{pratt1,maksin,sinyu,prattcso}.

\section{EXPERIMENT}
The CERES/NA45 spectrometer at the CERN SPS is optimized for the
measurement of low-mass electron pairs in the pseudorapidity
range 2.1$<$$\eta$$<2.65$. It consists of two Silicon
Drift Detectors (SDDs), located about 12~cm downstream of the segmented
Au-target, and two Ring Imaging Cherenkov counters for electron identification. 
To improve the mass resolution,
the experimental setup was upgraded in 1998 by the addition of a cylindrical
Time Projection Chamber (TPC) with radial electric drift field~\cite{anaqm}. 
The TPC is located behind the existing spectrometer, 3.8~m 
downstream of the target.
It is operated 
inside the field of a new magnet system.
Analysis of the track curvature leads to a determination of
the momentum of charged particles.
The measurement of up to 20 space points per charged particle track
allows the determination of the specific energy loss d$E$/d$x$ with a 
precision of 10.5\%, thereby supplementing the particle identification
capability of the spectrometer.
In addition to improving the invariant mass resolution of
electron pairs, the TPC also adds sensitivity to hadronic observables
close to midrapidity~\cite{harryqm,wollisqm}.

\section{DATA ANALYSIS}
In this paper we present the results of a two-pion interferometry
analysis in 40, 80, and 158 AGeV Pb+Au collisions~\cite{heinzphd}. 
The data were taken
during the 1999 (40 AGeV) and 2000 (80 and 158 AGeV) Pb-beam periods. 
In 1999, the segmented Au-target consisted of 8 subtargets,
separated by 3.1~mm in beam direction, 
and a thickness of 25$\mu$ each,
adding to a total hadronic interaction length of 0.82\%. 
For the 2000 Pb-beam period, 
the target
system was replaced by 13 subtargets of the same thickness and
2~mm spacing, with a total interaction length of 1.33\%.
An online centrality selection of approximately the upper
30\% of the total geometric cross section was applied,
using the pulse height deposited by charged particles in the
SDDs (in 1999) or in a Scintillator Multiplicity Counter (in 2000).
The same detector information is used for an
offline characterization of the centrality of the events
measured in the 2000 run period,
while for the 1999 data set the number of reconstructed charged
particle tracks in the SDDs is used.
Contributions from non-target interactions are found to be
negligible.
The data are presented in four bins of centrality, and 
the corresponding
average numbers of participants are derived from a geometric
nuclear overlap model~\cite{eskola} with $\sigma_{\rm NN}$=30~mb,
resulting in a total cross section of $\sigma_G$=6.94~barn. 
Information about the four
centrality bins is collected in Table~\ref{tab:centralities}.

\begin{table}[h!]
\centering
\vglue0.2cm
\caption{\label{tab:centralities}Definition of centrality bins.}
\vspace{0.25cm}
\begin{tabular}{|c|c|c|c|} 
\hline
      Centrality bin  \rule[-2mm]{0mm}{6mm} & $\sigma / \sigma_{\rm geo}$ & $b_{\rm min} - b_{\rm max}$ &
      $\left\langle N_{\rm part}\right\rangle$ \\ \hline \hline

      1 \rule[-2mm]{0mm}{6mm} & $>$15\%  & 5.8-6.5 & 202 \\ \hline
      2 \rule[-2mm]{0mm}{6mm} & 10-15\%  & 4.7-5.8 & 236 \\ \hline
      3 \rule[-2mm]{0mm}{6mm} &  5-10\%  & 3.3-4.7 & 287 \\ \hline
      4 \rule[-2mm]{0mm}{6mm} & $<$ 5\%  & $<$ 3.3 & 347 \\ \hline

\end{tabular}
\end{table}

\begin{figure}[h!]
   \centering
   \includegraphics[width=15cm]{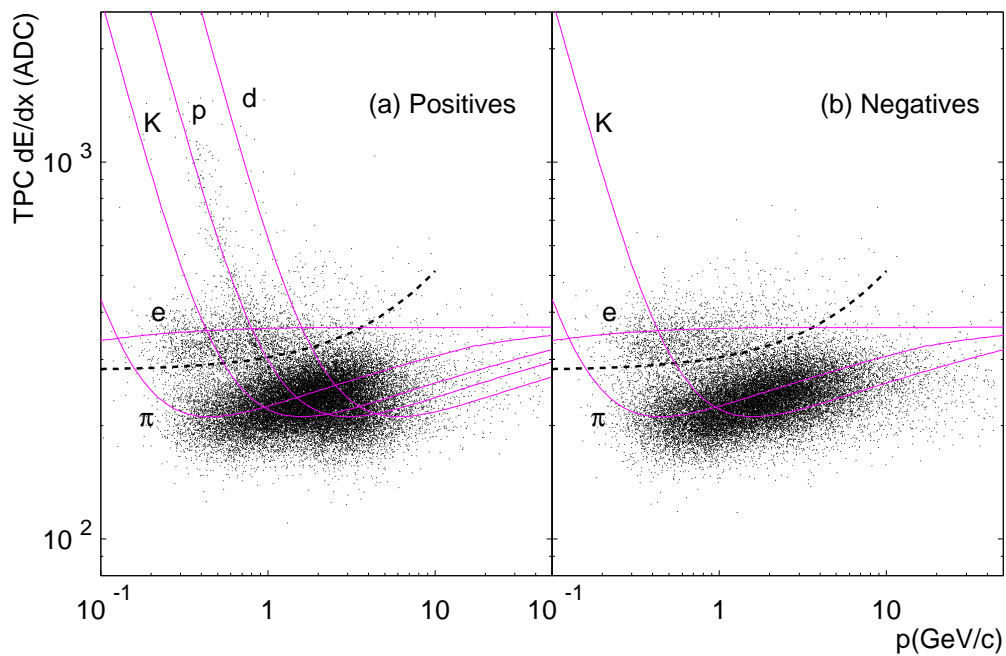}\\
   \caption{\label{dedx}The specific energy loss d$E$/d$x$ in the CERES-TPC in units of ADC 
                counts for positive (a) and negative (b) particles 
                as function of the particle momentum. 
                The cuts used to enrich the pion sample are indicated
                by the dashed lines. Shown as solid lines 
                are parametrizations~\cite{lars} of the specific energy loss 
                for different particle
                species according to the 
                Bethe-Bloch formula.}
\end{figure}

For the determination of the correlation function we combine like-sign
pairs of charged particles detected in the TPC. A momentum dependent
cut on the specific energy loss d$E$/d$x$ in the TPC
cleans the track sample from soft electrons and protons (see Fig.~\ref{dedx}).
Kaons are removed only for momenta below 0.5~GeV/c.
Requiring a match
of the TPC track to the SDD system suppresses the contribution 
from secondary decay products and results in an increase of the
correlation strength $\lambda$ but does not alter the extracted source
radii.

Similarly, taking the direction of the track from the SDDs rather than from the TPC,
which reduces the effect of multiple scattering, leaves the extracted source 
radii unchanged.  
Since SDD information is not available for the full data sample we 
used SDD information only for the tests just mentioned.
For the results presented later we restricted
the analysis to TPC information only.
At the present stage of the spectrometer calibration
the momentum resolution is $\delta p/p$=3-5\% in the momentum range
of interest, 0.25$<p<$3~GeV/c, 
consistent with the observed widths of reconstructed 
neutral strange hadrons $\Lambda$ and $K^{\circ}_{s}$~\cite{heinzphd}. 

For each particle pair we calculate the mean transverse momentum 
\begin{equation}
k_{t}=\frac{1}{2}|\vec{p}_{t,1}+\vec{p}_{t,2}|
\end{equation}
and the pair rapidity 
\begin{equation}
y_{\pi\pi}=\frac{1}{2}\ln\frac{E_{1}+E_{2}+p_{z,1}+p_{z,2}}
{E_{1}+E_{2}-p_{z,1}-p_{z,2}}. 
\end{equation}
The rapidity 
$y_{\rm mid}$ (midrapidity)
of the c.m.~system of the collision 
changes as function
of the beam energy from $y_{\rm mid}$=2.23 at 40~AGeV to $y_{\rm mid}$=2.91
at 158~AGeV, while the acceptance of the spectrometer is fixed.
For the determination of the correlation 
functions we restrict the analysis to pairs with $k_{t}>0.05$~GeV/c and
-0.25$<y_{\pi\pi}$-$y_{\rm mid}<$0.25 (40~AGeV), -0.5$<y_{\pi\pi}$-$y_{\rm mid}<$0 (80~AGeV), and 
-1$<y_{\pi\pi}$-$y_{\rm mid}<$-0.5 (158~AGeV).
We split the three-momentum difference vector $\vec{q}$ of two like-sign particles
into components, $\vec{q}$=$(q_{\rm long},q_{\rm side},q_{\rm out})$, 
to obtain detailed information about the space-time evolution.
Following Bertsch and Pratt~\cite{ber,pra}, $q_{\rm long}$ is the momentum difference along
the beam direction, calculated in the longitudinal rest frame (LCMS) of the pair,
$q_{\rm out}$ is parallel to the pair transverse momentum vector $\vec{k_{t}}$ and 
$q_{\rm side}$ is perpendicular to $q_{\rm long}$ and $q_{\rm out}$. 
Note that the pion mass is assumed for the calculation of the LCMS.

The correlation function is defined as the ratio 
$C_{2}(\vec{q})$=$A_{2}(\vec{q})/B_{2}(\vec{q})$
where the `signal' $A_{2}(\vec{q})$ is the probability to find a pair with momentum difference
$\vec{q}$ in a given event 
and the `background' $B_{2}(\vec{q})$ is the corresponding mixed-event distribution.
For the construction of $B_{2}(\vec{q})$, only tracks from events of the same 
centrality bin are combined.
The number of pairs in $B_{2}(\vec{q})$ was chosen to be 
about ten times larger than
in $A_{2}(\vec{q})$ to keep the statistical error of the background negligible. 
The normalization is part of the fitting procedure (see below).

A separate analysis of positive and negative like-sign 
pairs gave within statistical errors the same results for the 
extracted source radii. To improve statistics, 
we have thus combined 
the results of positive and negative like-sign pairs by adding the signal and the 
background distributions, 
respectively, before the construction of $C_{2}(\vec{q})$.
This procedure is appropriate if the normalization constants of the positive and
negative pair samples are identical, which has been verified.
 
\begin{figure}
   \centering
   \includegraphics[width=12cm]{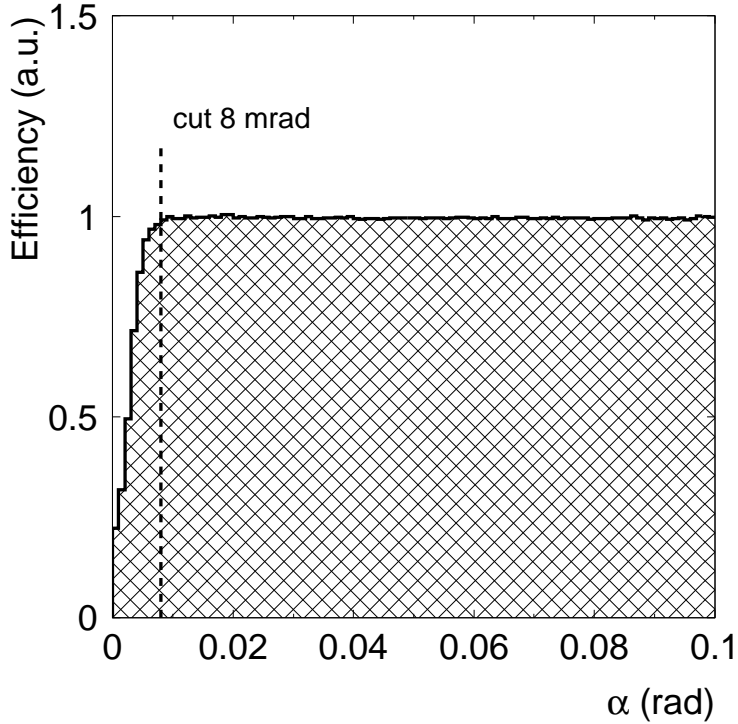}\\
   \caption{Relative two track reconstruction efficiency as function of
        the opening angle~$\alpha$. For more details see text.}
   \label{ttreso}
\end{figure}
Due to the correlation between the momentum difference of a pair
and its spatial distance in the detector, the
finite two track separation power of the TPC can lead to a systematic 
bias of the correlation function.
Fig.~\ref{ttreso} shows the opening angle distribution of like-sign
particle pairs divided by the corresponding mixed-event 
distribution. 
The resulting pair reconstruction efficiency has been normalized
to unity at large opening angles.
The pair efficiency is independent
of the opening angle $\alpha$ except for very small $\alpha$.
We omitted 
pairs with opening angles $\alpha$$<$8~mrad both in the construction
of $A_{2}(\vec{q})$
and $B_{2}(\vec{q})$
to avoid a distortion of the correlation function due to
the finite two-track separation power. 
This opening angle corresponds to a spatial distance of
about 3~cm at the entrance of the TPC.
Our studies have shown that the extracted source radii change only
within their statistical errors if the two-track separation cut
is varied between 6 and 10~mrad.

\begin{figure}
   \centering
   \includegraphics[width=12cm]{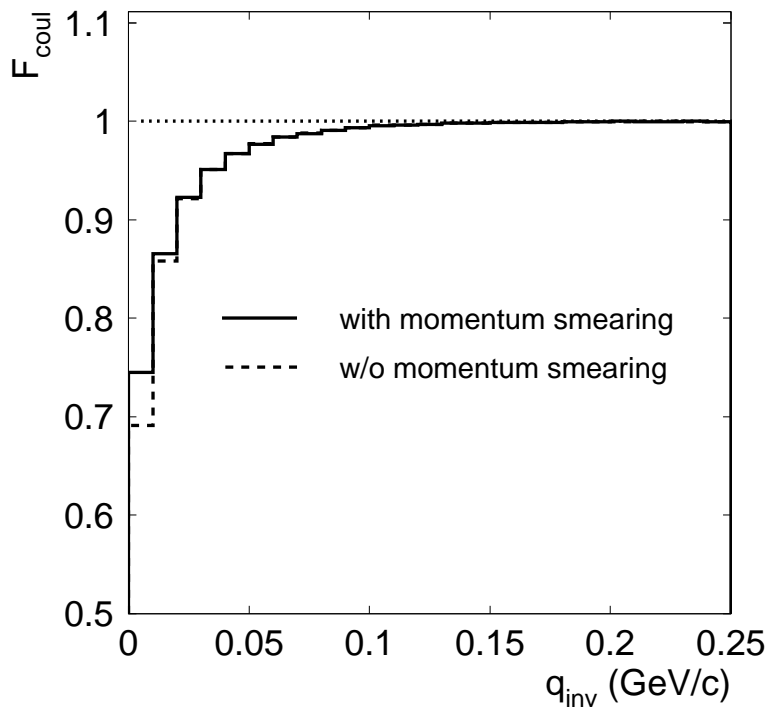}\\
   \caption{Coulomb correlation function $F_{\rm coul}$ as function of $q_{\rm inv}$ 
        before and after smearing with the momentum resolution.}
   \label{coulomb}
\end{figure}

Correlations in momentum space are caused not only by quantum statistics but also 
by final state interactions, the most important one being the mutual
Coulomb repulsion. 
Since this is a two-body process, it depends only on the relative momentum 
of the pair in its rest frame 
$q_{\rm inv}$=$\sqrt{|(\vec{p_1}-\vec{p_2})^2-(E_1-E_2)^2|}$.
To account for this, we parametrized the $q_{\rm inv}$-dependent 
Coulomb correlation function
$F_{\rm coul}(q_{\rm inv})$ 
derived by integration of the Coulomb wave functions
over a Gaussian source of 5~fm size~\cite{baympbm}.
%as calculated by
%a semi-classical approach assuming a source size of 5~fm~\cite{baympbm}.
In the next step, the impact of the finite detector resolution of 
$\sigma(q_{\rm inv})$=5-10~MeV/c (see below)
on $F_{\rm coul}(q_{\rm inv})$ was evaluated using a Monte-Carlo simulation. 
The resulting Coulomb correlation functions $F_{\rm coul}(q_{\rm inv})$ before and after
momentum smearing are shown in Fig.~\ref{coulomb}.

In most of the previous HBT analyses the Coulomb correction was carried 
out by applying
$q_{\rm inv}$-dependent weights to each pair in the background distribution
$B_{2}(\vec{q})$. Afterwards, the Coulomb-corrected correlation function
$C_{2}(\vec{q})$=$A_{2}(\vec{q})/B_{\rm 2,corr}(\vec{q})$ was 
normalized in the region of large $\vec{q}$ and fit by a Gaussian
parametrization, which then should describe the pure BE correlation peak.
This approach assumes that all particle pairs in the signal distribution
are subject to the Coulomb correlation, which is in general 
not true under realistic physical and experimental conditions.
In the following, we argue that physics and detector related effects
lead to a reduction of the observed BE correlation strength, 
usually expressed in terms of the parameter $\lambda$,
and that the same effects 
also lead to a similar reduction of the Coulomb repulsion. 
Therefore,
the strength of the Coulomb correction applied to the data should
be linked to the experimentally observed $\lambda$ parameter.

A reduction of the observed BE correlation from its 
theoretical value of two at vanishing relative momentum can
be caused by coherent pion production, pions from
long-lived resonances and weak decays, particle misidentification,
and finite momentum resolution. 
We assume that the contribution from coherent pion production is
negligible because
the observed pion phase space densities at SPS energies~\cite{ferenc} are too low to expect
a significant degree of coherence of pion production. 
Pions from long-lived resonances such 
as the $\omega$(782) add a tail to the density
distribution of the pion source, which then results in a 
narrow spike of the correlation function at very small relative momentum.
Due to the finite
momentum resolution of the detector the spike cannot be resolved.
This leads to a reduction of the
observed $\lambda$ parameter. Since the effective source size of 
pions emerging from these resonances is large, their mutual Coulomb 
interaction can be assumed to be negligible. 
The same arguments apply for pions from weak decays of e.g.~$\Lambda$ and $K^{\circ}_{s}$.
A contamination of the track sample by particles other than
pions leads to pairs of like-sign non-identical particles. 
They obviously do
not contribute to the BE correlation, but they experience Coulomb repulsion. 
However, due to their different masses, the strongest Coulomb correlation
for those pairs does in general not occur at $\vec{q}$=0 in the LCMS,
which is calculated assuming the pion mass, but at finite $\vec{q}$ 
away from the pion BE peak. 
We have evaluated this for the CERES acceptance and $K\pi$ and $p\pi$ pairs.
In any case, 
we conclude that the contribution to Coulomb
correlation of non-identical pairs is weak,
since the measured correlation functions 
are flat around the corresponding $q_{\rm inv}$.
Note that pairs containing
electrons or positrons do not contribute to Coulomb correlations because
most of them stem from decays or conversions far away from the interaction
region.
Pairs of identical particles which are not pions can give rise to BE and
Coulomb correlations. However, a detailed comparison of the correlation functions
of positive and negative like-sign pairs showed no significant
difference of the
extracted source radii and of the shape of the correlation functions
at large $\vec{q}$, although the level of contamination by non-pion
pairs is very different in the positive and negative pair sample
(see Fig.~\ref{dedx}).
Therefore we conclude that the contribution of non-pion pairs to 
Coulomb repulsion and BE correlation is negligible. 

In essence, we argue that only those pairs which contribute to
the measured BE correlation strength also contribute to 
Coulomb repulsion. 
Therefore, the applied Coulomb correction should
be scaled according to the observed $\lambda$ parameter.
Since the value of the $\lambda$ parameter is {\em a priori}
unknown, we construct the
correlation function from the signal and
background distributions without Coulomb correction, 
and include the effect of Coulomb repulsion
into the fit function:
\begin{eqnarray}
  C_{2}(\vec{q})&=&A_2(\vec{q})/B_2(\vec{q}) = N\cdot\left[1+\lambda^{'}\left((1+G)\cdot F^{\star}-1\right)\right], 
  \label{c2para}
\end{eqnarray}
with
\vspace{-0.5cm}
\begin{eqnarray}
        G       &=& \exp(-R_{\rm long}^{2}q_{\rm long}^{2}
                                    -R_{\rm side}^{2}q_{\rm side}^{2}  \nonumber \\
        & &                      -R_{\rm out}^{2}q_{\rm out}^{2}
                                    -2R_{\rm out,long}^{2}q_{\rm out}q_{\rm long}),~~~~{\rm and} \\
  	F^{\star}&=& w(k_{t})\cdot (F_{\rm coul}(q_{\rm inv})-1)+1.
        \label{fstar}
\end{eqnarray}
The fit function contains the usual Gaussian parametrization $G$, with
the Gaussian source radii $R_{\rm long}$, $R_{\rm side}$, and $R_{\rm out}$,
and the cross-term $R_{\rm out,long}^{2}$ which appears as a consequence of space-time
correlations in non-boost-invariant systems~\cite{heinzrol}.
The normalization $N$ takes into account the different statistics used for the
construction of $A_{2}(\vec{q})$ and $B_{2}(\vec{q})$.
The Coulomb term $F^{\star}$ contains the two-pion Coulomb correlation function
$F_{\rm coul}(q_{\rm inv})$ which was derived as described before.
Its application requires supplementary information about the average
$q_{\rm inv}$ in each of the 10~MeV/c $(q_{\rm long},q_{\rm side},q_{\rm out})$-bins,
which was stored in separate arrays of the same bin size.
Following the above argumentation,
the strength of the Coulomb correction needs to be adjusted to the
$\lambda$ parameter.
At this point, an additional complication arises due to
the finite momentum resolution, which leads to a $k_t$ dependent reduction of 
$\lambda$ by 13\% to 45\%, i.e.~$\lambda$=$\lambda^{'}$$\cdot$$w(k_{t})$. 
The $k_t$ dependent correction 
factors $w(k_{t})$ were determined by a Monte Carlo simulation
and range from 1.15 in the lowest $k_{t}$-bin to 1.82 in the highest $k_{t}$-bin.
The inclusion of $w(k_{t})$ into the Coulomb term $F^{\star}$ 
accounts for the apparent depletion of $\lambda$ due to the finite momentum 
resolution and assures the appropriate scaling of the Coulomb correction. 
We note that Eq.~(\ref{c2para}) is identical to the parametrization
proposed in~\cite{sinyuled} for perfect resolution, e.g.~$w(k_{t})$=1.

\begin{figure}
   \centering
   \includegraphics[width=9.8cm]{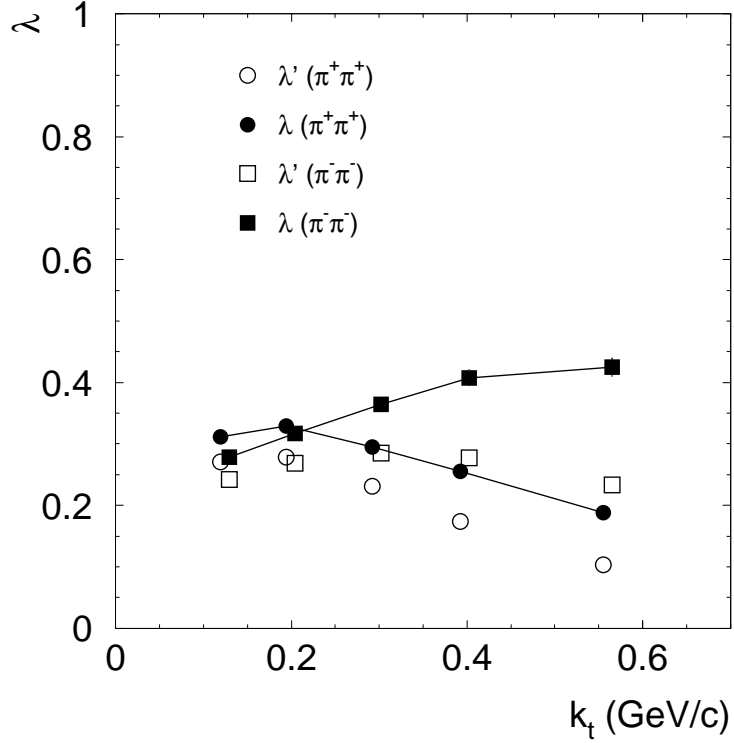}\\
   \caption{Parameters $\lambda^{'}$ (open symbols)
        and $\lambda=w(k_t)\cdot \lambda^{'}$ (full symbols)
        as function of $k_t$ obtained from an analysis of 
	$\pi^+\pi^+$ and $\pi^-\pi^-$ pairs
        in central 158 AGeV Pb+Au collisions. For explanation see text.}
   \label{lambda}
\end{figure}
In Fig.~\ref{lambda} are shown the parameters $\lambda^{'}$ 
and $\lambda$=$\lambda^{'}$$\cdot$$w(k_{t})$
for $\pi^+\pi^+$ and $\pi^-\pi^-$ correlations at 158 AGeV ($\sigma/\sigma_{\rm geo}$$<$30\%).
The $\lambda$ parameter for negative pions 
increases with $k_{t}$ as expected from a decreasing
contribution from electrons and pions from resonance decays.
In contrast, for positive pions
the $\lambda$ parameter decreases with $k_{t}$ because
of an increasing contamination by protons. This behaviour 
is in quantitative agreement with the contribution
from misidentified particles and pions from resonance 
decays as seen in Monte-Carlo simulations.

\begin{figure}
   \centering
   \includegraphics[width=15cm]{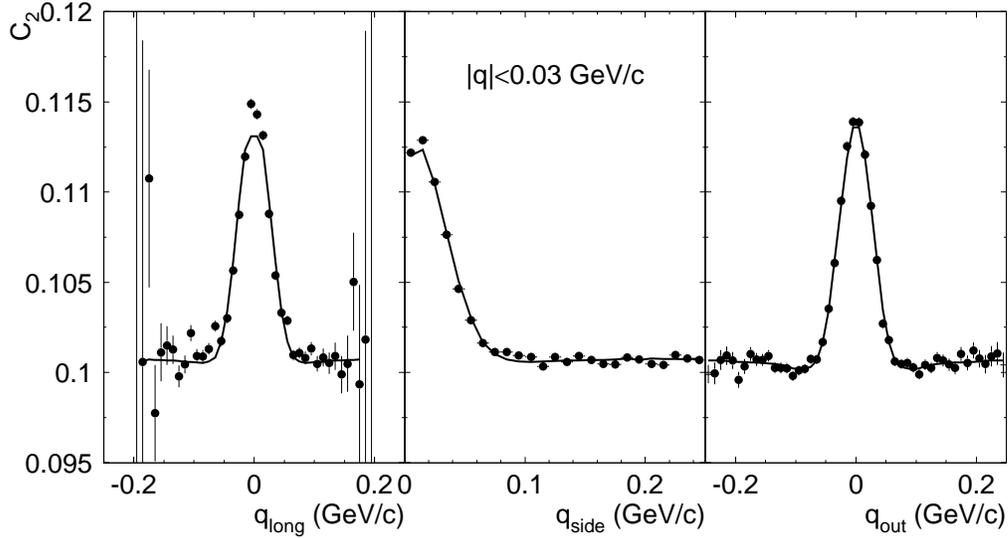}\\
   \caption{Projections of the two-pion correlation function in central 158 AGeV Pb+Au
            collisions (0.15$<$$k_{t}$$<$0.25 GeV/c). The data are integrated over
            the range $|q|$$<$0.03~GeV/c in the non-projected coordinates. 
        Also shown are the results from 
            a fit using Eq.~(\ref{c2para}).}
   \label{projs}
\end{figure}

In Fig.~\ref{projs} we present examples of one-dimensional projections
of the correlation function, integrated over the range $|q|$$<$30~MeV/c
in the other two components. 
Also shown are the corresponding
projections of the fit function Eq.~(\ref{c2para}). 
\begin{figure}
   \centering
   \includegraphics[width=15cm]{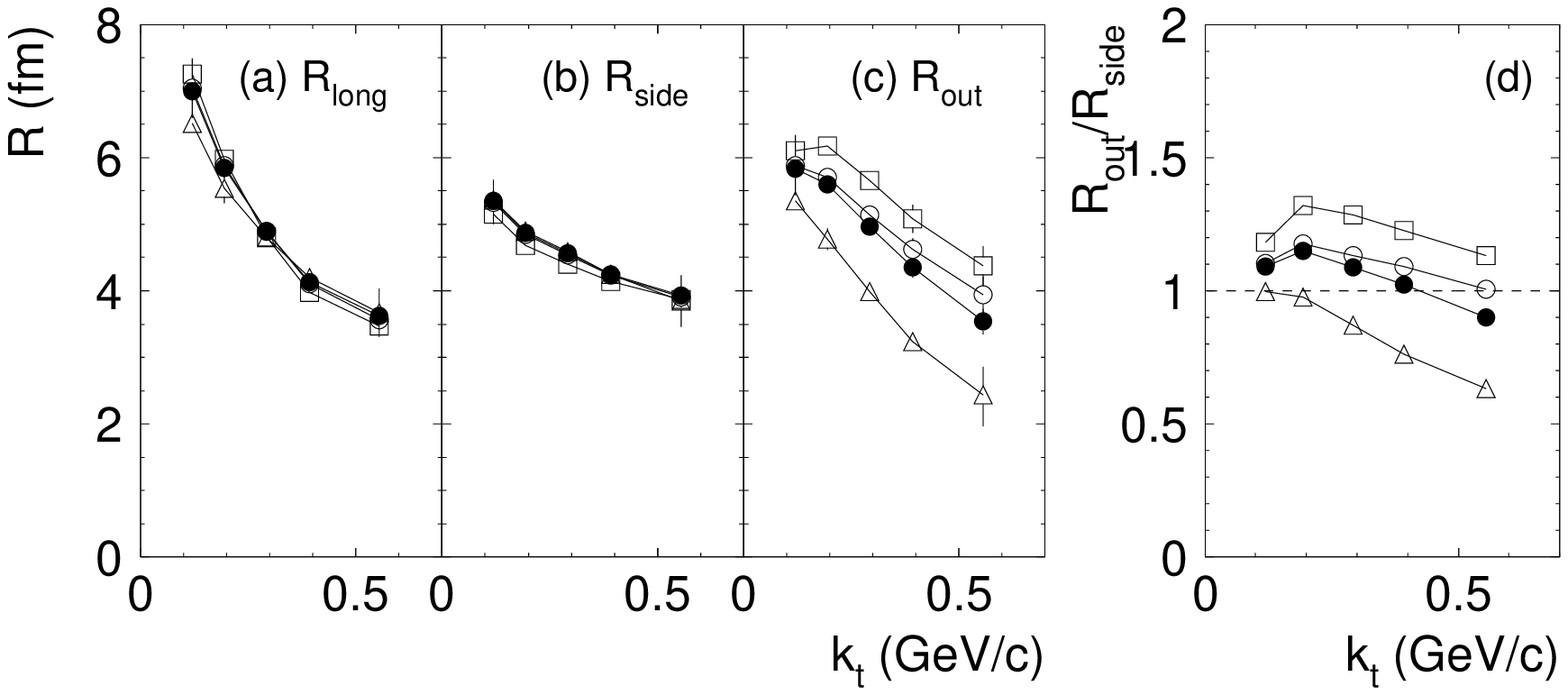}\\
   \caption{Radius parameters (panels (a-c)) and ratio 
	$R_{\rm out}$/$R_{\rm side}$ (panel (d))
        as function of $k_t$ assuming different strength of the Coulomb 
	repulsion (see text). 
	Shown are the results for (1.) full Coulomb strength (open 
	triangles), (2.) no Coulomb repulsion (open squares),
	(3.) Coulomb repulsion coupled to the $\lambda$ parameter as in 
	Eq.~(\ref{c2para}) (full circles), and (4.) same as (3.) 
	but with $w(k_t)=1$ (open circles).}
   \label{coulsyst}
\end{figure}
We want to emphasize 
that the depletion of the correlation function due to Coulomb repulsion
is not only visible at the smallest $q$ but also in the region outside
the BE correlation peak. 
For kinematical reasons this effect is most pronounced in $q_{\rm out}$ 
and increases with $k_{t}$. 
An incorrect treatment of the Coulomb
repulsion strength therefore affects $R_{\rm side}$ and $R_{\rm out}$
in a different way, in particular leading to a $k_{t}$-dependent 
bias of the $R_{\rm out}$/$R_{\rm side}$ ratio.
This is demonstrated in Fig.~\ref{coulsyst}, where the source 
radii and the ratio $R_{\rm out}$/$R_{\rm side}$ 
obtained for different Coulomb strengths from central 158 AGeV events
are shown.
We compare the results assuming full Coulomb strength (open triangles), where all 
background pairs were corrected by $F_{\rm coul}(q_{\rm inv})$, with
different implementations of the Coulomb term $F^{\star}$ in Eq.~(\ref{c2para}). 
We considered the cases $F^{\star}$=1, corresponding to no Coulomb 
correction (open squares), $F^{\star}$=$F_{\rm coul}(q_{\rm inv})$
(open circles) as proposed in~\cite{sinyuled}, and finally 
$F^{\star}$=$w(k_{t})\cdot (F_{\rm coul}(q_{\rm inv})$-1)+1, as given
in Eq.~(\ref{fstar}) (full circles).
The parameters $R_{\rm long}$ and $R_{\rm side}$ depend very little on the
assumed strength of the Coulomb repulsion, however, the results for $R_{\rm out}$ and 
correspondingly $R_{\rm out}$/$R_{\rm side}$ are very 
sensitive to the procedure employed~\footnote{Analysis of the 
RHIC HBT data~\cite{starhbt} 
using the procedure described above should lead to an increase 
of $R_{\rm out}$ and $R_{\rm out}$/$R_{\rm side}$~\cite{frankqm}. 
Whether this
can explain part or all of the RHIC HBT `puzzle' remains to be investigated.}.
We also note that the inclusion of $w(k_{t})$ has only little effect 
on the extracted source radii.
In this context it should be mentioned that our fit function 
Eqs.~(\ref{c2para})-(\ref{fstar})
describes very well not only the Gaussian BE correlation peak
but also the observed undershoot of the baseline at large $q_{\rm out}$
(see Fig.~\ref{projs}, right panel).

\begin{figure}
   \centering
   \includegraphics[width=12cm]{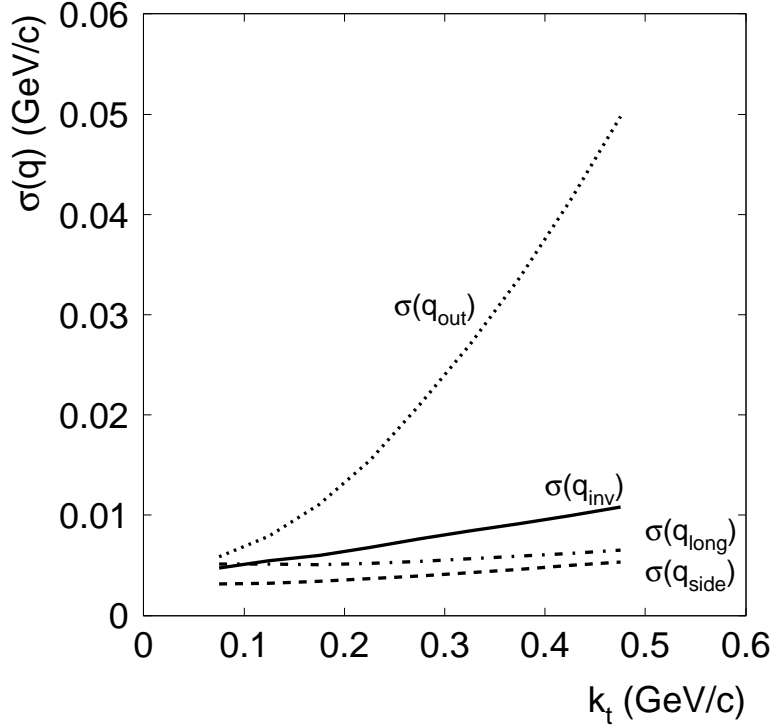}\\
   \caption{Resolution in $q_{\rm inv}$, $q_{\rm long}$, $q_{\rm side}$, and $q_{\rm out}$ 
           as determined by a Monte Carlo simulation.}
   \label{qreso}
\end{figure}
\begin{figure}
   \centering
   \includegraphics[width=15.5cm]{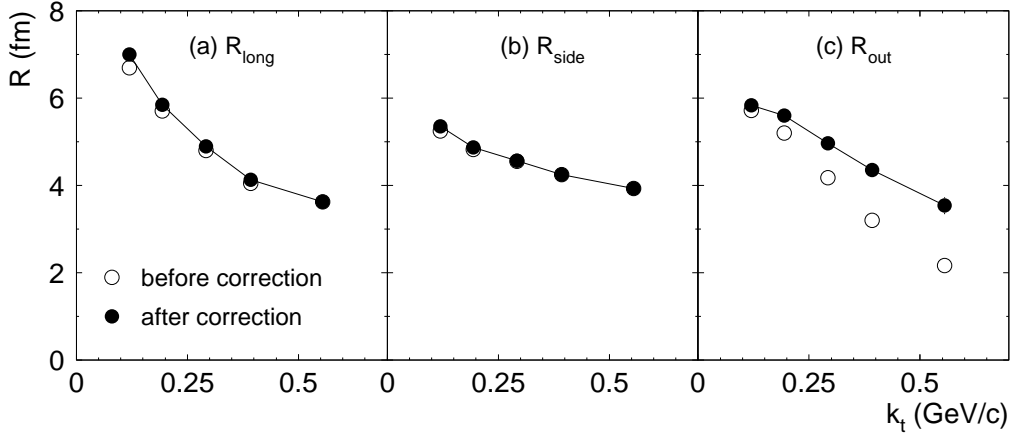}\\
   \caption{Left to right: The radius parameters $R_{\rm long}$, $R_{\rm side}$, and
        $R_{\rm out}$ as function of $k_t$. The open symbols are before, the full symbols
        are after correction for the finite momentum resolution.}
   \label{momcorr}
\end{figure}

The finite angular and momentum resolution of the TPC 
cause a systematic reduction of the extracted source radii. 
Corrections were determined by a Monte-Carlo simulation
and depend on $k_{t}$ as well as 
on the extracted source radii themselves.
The resolution in the components of $\vec{q}$ is shown in Fig.~\ref{qreso}.
Since $\sigma(q_{\rm long})$ and $\sigma(q_{\rm side})$
are 3-7~MeV/c only, the corrections to be applied to $R_{\rm long}$ and 
$R_{\rm side}$ are below 2\%. 
This is demonstrated in Fig.~\ref{momcorr} where the source radii from central 158 AGeV events
before and after correction for the finite momentum resolution are shown.
However, $\sigma(q_{\rm out})$ deteriorates significantly as function of $k_{t}$,
which implies corrections up to 60\% at high $k_{t}$ and consequently 
larger systematic uncertainties in $R_{\rm out}$ at these momenta. 

We conclude that the main sources of systematic errors are the remaining uncertainties
in the treatment of the Coulomb repulsion and the corrections for the finite momentum
resolution. Varying the Coulomb strength and the momentum resolution within realistic
limits, we estimated the systematic errors in $R_{\rm long}$ and
$R_{\rm side}$ to be 4\%. In $R_{\rm out}$, the systematic error is larger and
depends on $k_t$. We estimated that the systematic error in $R_{\rm out}$ is 5\%
in the lowest and 20\% in the highest $k_t$-bin. In the following we present the
results with only statistical errors at the data points. The systematic errors are
indicated by shaded bands in the figures.

\section{RESULTS}

\begin{figure}
   \centering
   \includegraphics[width=15cm]{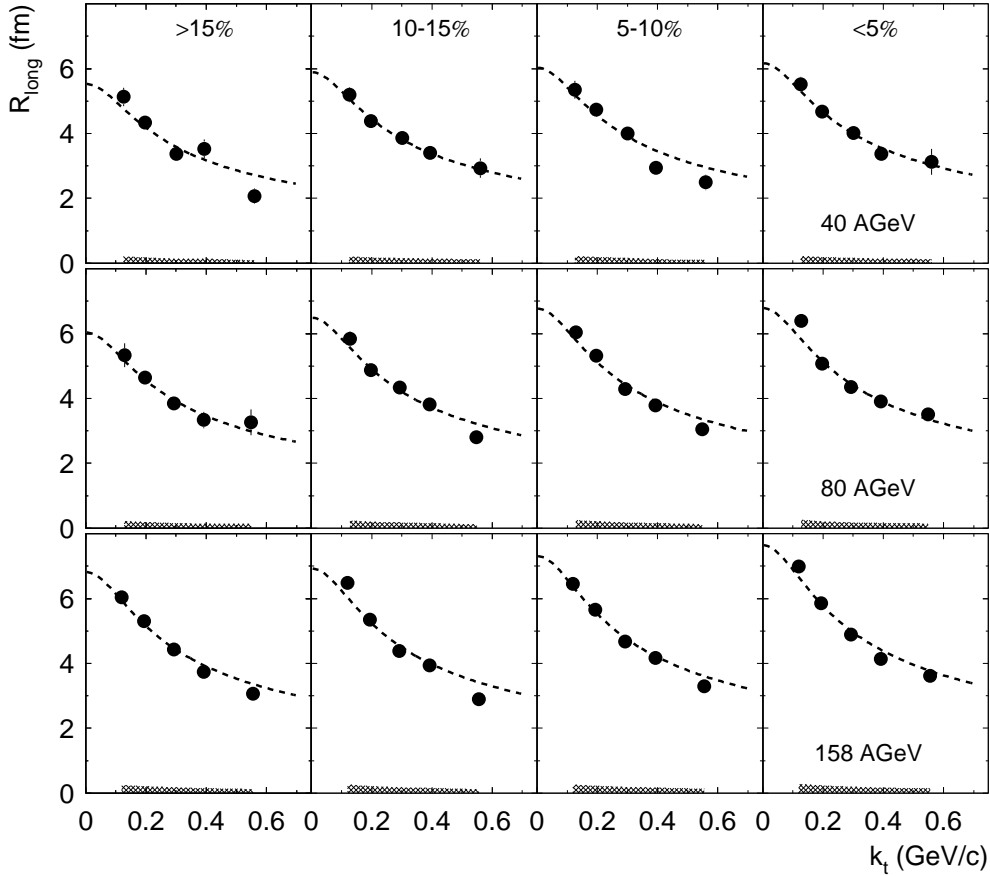}\\
   \caption{The longitudinal radius parameter $R_{\rm long}$ as function 
                of $k_{t}$ in different bins of centrality at 40, 80, and
                158 AGeV. Also shown as dashed lines are fits to the data 
                assuming longitudinal boost invariance (see text).
                The shaded regions indicate the systematic error.}
   \label{rlong}
\end{figure}
\begin{figure}
   \centering
   \includegraphics[width=14cm]{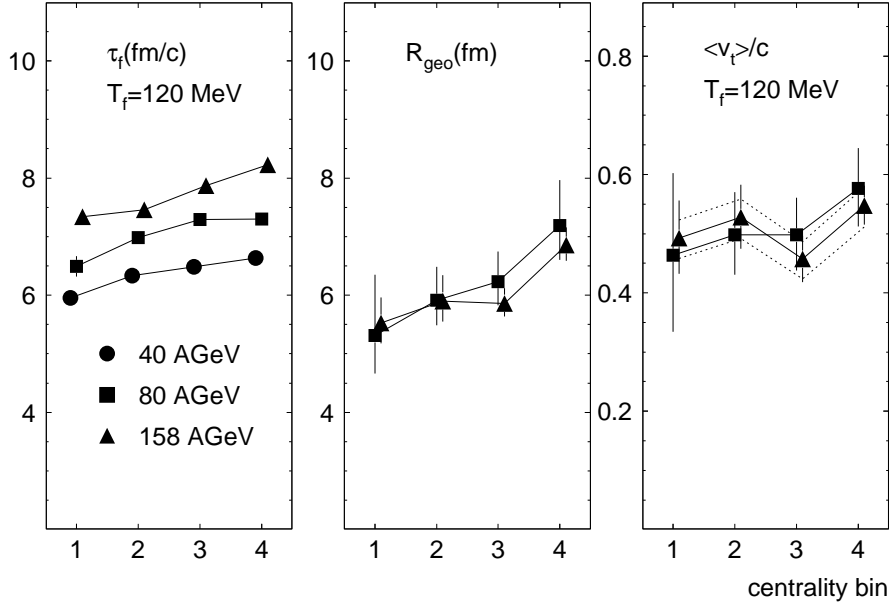}\\
   \caption{Centrality and beam energy dependence of the duration
        $\tau_{\rm f}$ of expansion, the geometric transverse source size $R_{\rm geo}$
        and the average transverse flow velocity $\langle v_{t} \rangle$ derived
        from fits to the $k_{t}$-dependence of $R_{\rm long}$ and $R_{\rm side}$
	(see text).
        The dashed lines in the right panel indicate the results obtained 
        at 158 AGeV
        for a change of the freeze-out temperature $T_f$ by $\pm$20~MeV.}
   \label{fo_parms}
\end{figure}

We present, in Fig.~\ref{rlong}, the results for the longitudinal 
source radius $R_{\rm long}$ 
as function of $k_{t}$ 
and in different bins of centrality for all three beam energies. 
At all beam energies and centralities under investigation
the $R_{\rm long}$ values exhibit a very pronounced $k_{t}$-dependence. This is characteristic for the case
of a strong longitudinal expansion, where the length of homogeneity is
entirely saturated by the thermal velocity scale $\sqrt{T_{f}/m_{t}}$~\cite{maksin,sinyu}. 
For the limiting case of a boost-invariant expansion in longitudinal direction the
following relation was given by Sinyukov~\cite{sinyu}, which allows to extract the
duration $\tau_{f}$ of the expansion (the `lifetime') of the system: 
\begin{equation}
R_{\rm long}=\tau_{f}(T_{f}/m_{t})^{\frac{1}{2}}. 
\label{sinyuform}
\end{equation}
In real sources boost-invariance is violated by the finite extension
of the system in longitudinal direction. Sophisticated model analyses
which consider more realistic density profiles 
do, however, extract lifetimes which are quite similar to those obtained
from Eq.~(\ref{sinyuform}) (see e.g.~\cite{borisana}).
Fitting expression~(\ref{sinyuform}) to the data
and assuming a constant thermal freeze-out temperature $T_{f}$=120~MeV results in a smooth
increase of $\tau_{f}$ both with centrality and beam energy 
from about 6 to 8~fm/c as shown in the left panel of Fig.~\ref{fo_parms} and 
Table~\ref{tab:fitresults}.
The choice of $T_{f}$=120~MeV is motivated by previous analyses 
of AGS and SPS data~\cite{pbmqm97,na49hbt}, which lead to a good description of single 
particle spectra and HBT correlations. Since the temperature enters only 
to the power $1/2$, unrealistically large changes in $T_{f}$ are required 
to significantly
affect $\tau_{f}$. We note that the 
increase of $R_{\rm long}$ with centrality and beam energy could alternatively
be explained by a constant duration of expansion $\tau_{f}$, if
$T_{f}$ increases by 25\% with centrality and by 50\% 
from the lowest to the highest beam energy.

\begin{figure}
   \centering
   \includegraphics[width=15cm]{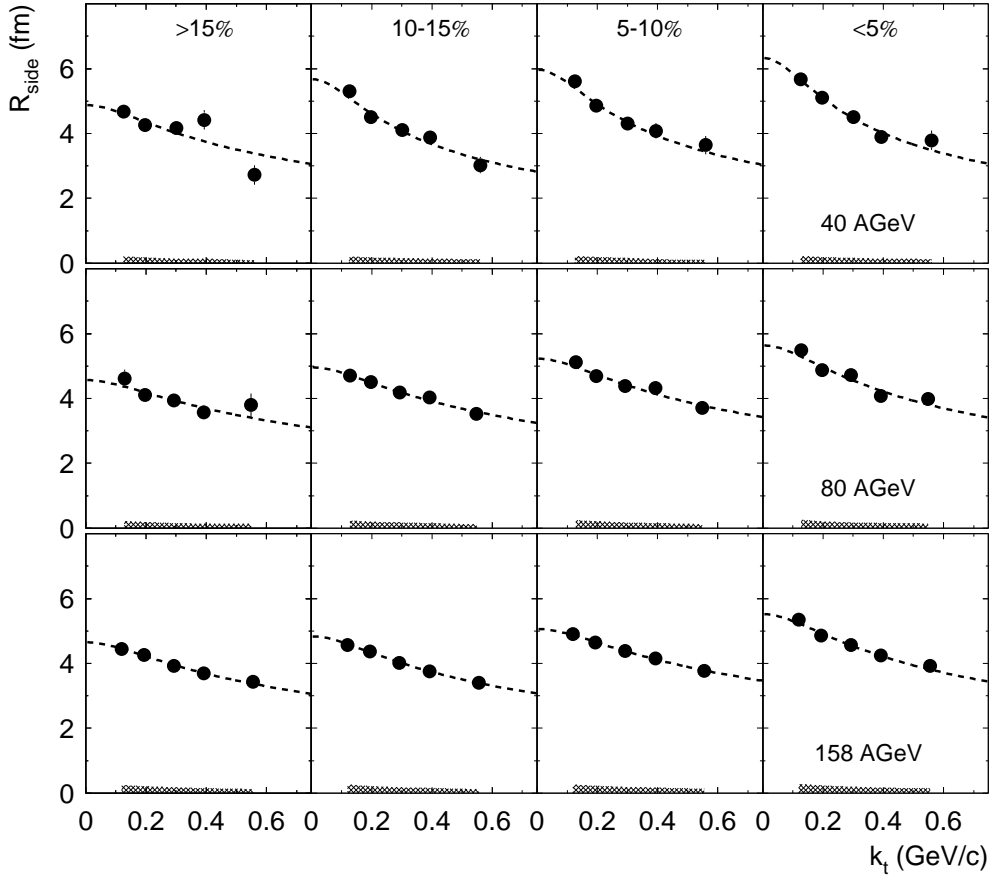}\\
   \caption{The sideward radius parameter $R_{\rm side}$ as function
                of $k_{t}$ in different bins of centrality at 40, 80, and
                158 AGeV. Also shown are fits to the data (see text).
                The shaded regions indicate the systematic error.}
   \label{rside}
\end{figure}

The results for $R_{\rm side}$ are shown in Fig.~\ref{rside}.
They exhibit a weaker but still significant $k_{t}$-dependence,
indicating collective expansion also in transverse direction.
As expected from collision geometry, $R_{\rm side}$ slightly
increases with centrality. 
Comparing different beam energies, we observe that, at 40~AGeV, $R_{\rm side}$ at 
small $k_{t}$ is larger and the $k_{t}$-dependence is
systematically steeper than at the higher energies,
while the results at 80~and 158~AGeV are very similar to each other.
All measured $R_{\rm side}$ parameters at small $k_{t}$ are larger
by almost a factor of two compared to
the geometrical size of the initial overlap region of the two
colliding nuclei.

\begin{figure}
   \centering
   \includegraphics[width=15cm]{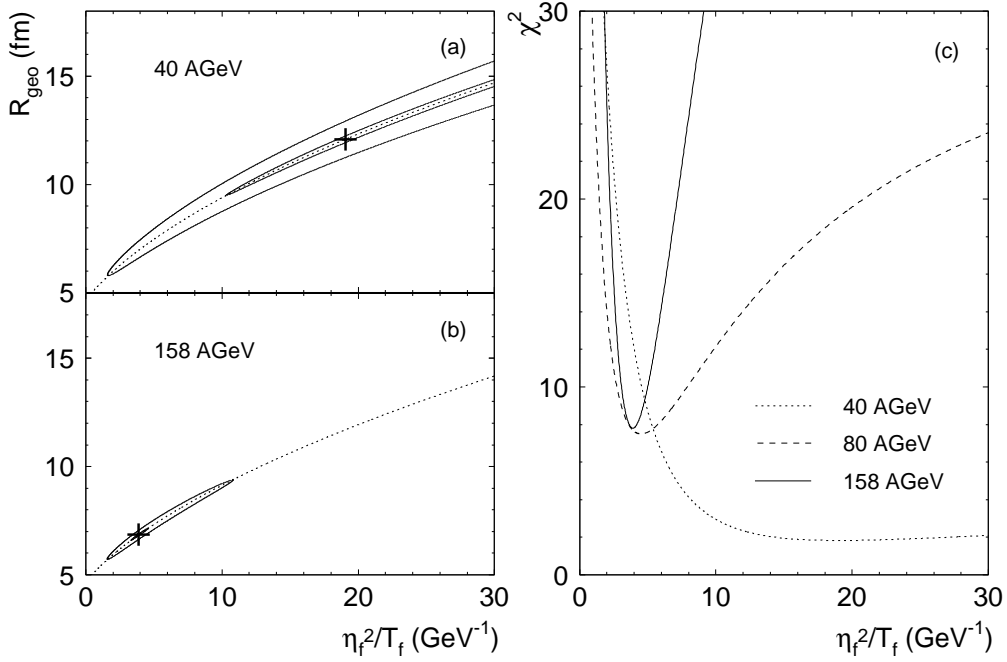}\\
   \caption{Panels (a-b) show the 1$\sigma$- and 30$\sigma$- contours
        of the fit of Eq.~(\ref{wiede}) to $R_{\rm side}$($k_t$) from 
        central 40 AGeV and 158 AGeV collisions in the
        $R_{\rm geo}$-$\eta_f^2/T_f$-plane. Also indicated 
	by the dotted lines is the location of
        $\chi^2$-valleys, and the position
        of the minimum by the crosses. In panel (c) are shown the $\chi^2$-values
        computed along the valleys indicated in (a-b) and plotted as function 
        of $\eta_f^2/T_f$.}
   \label{chi2}
\end{figure}

Hydrodynamically inspired model calculations predict an interplay between geometrical and
thermal length scales. Near midrapidity, $R_{\rm side}$ can be
approximated by~\cite{chapnix,csolo,scheibl}:
\begin{equation}
R_{\rm side}=R_{\rm geo}/(1+m_{t}\cdot\eta_{f}^{2}/T_{f})^{\frac{1}{2}}.
\label{wiede}
\end{equation}
This expression can be used to determine the `true' geometric transverse size $R_{\rm geo}$  
of the source at freeze-out. 
We estimate from a model dependent analysis 
of the fit parameter $\eta_{f}^{2}/T_{f}$ 
the average transverse flow velocity.
This fit parameter contains the thermal freeze-out temperature 
$T_f$ and the transverse
flow rapidity $\eta_{f}$, which characterizes the strength of transverse
flow for a given flow profile and transverse density distribution.
We applied a fit of this expression to the data, as indicated
by the dashed lines in Fig.~\ref{rside}. The fits describe the data very well
at all centralities and energies. 
However, we achieve
reasonable results and errors for the fit parameters only at 80 and 158 AGeV,
as shown in Table~\ref{tab:fitresults}. 
A closer inspection of the $\chi^2$-contours of the fits, shown in Fig.~\ref{chi2}, 
shows that the parameters
$R_{\rm geo}$ and $\eta_{f}^{2}/T_{f}$ are highly correlated and that 
the $\chi^2$-contours are very flat and without discernible minimum for the 40~AGeV data.
However, the $\chi^2$ values indicate that the 40~AGeV data are also consistent with 
fit parameters similar to those obtained for the higher energies, at in fact
the same level of confidence as at the higher energies.
We conclude that 
a separation between source geometry and flow dynamics
using expression Eq.~(\ref{wiede}) is not possible because 
of the steepness of $R_{\rm side}(k_{t})$ at
40 AGeV.

The results obtained at 80 and 158 AGeV are very similar to each other: We observe a smooth
increase of $R_{\rm geo}$ with centrality from about 5.5 to 7 fm, as shown in the
middle panel of Fig.~\ref{fo_parms}.
Assuming a thermal freeze-out temperature of $T_{f}$=120~MeV and a box-shaped
density distribution, 
the average transverse flow velocity $\langle v_{t} \rangle$
can be derived from the fit parameter $\eta_{f}^{2}/T_{f}$~\cite{wiedeqm,borisana}, 
as shown in 
the right panel of Fig.~\ref{fo_parms}. 
For the most central events at 80 and 158 AGeV we obtain 
$\langle v_{t} \rangle \approx 0.55 c$ 
and very little centrality dependence. 
A variation of the assumed freeze-out temperature $T_{f}$ by $\pm$20~MeV
alters the extracted flow velocities only within their statistical errors,
as demonstrated for the 158 AGeV results 
by the dashed band in Fig.~\ref{fo_parms} (right panel). 
These findings are similar to 
results from previous analyses~\cite{na49hbt,borisana,wa97hbt}.

\begin{table}[h!]
\begin{center}
\vglue0.2cm
\caption{Summary of freeze-out parameters obtained from fits to the $k_t$-dependencies
        of $R_{\rm long}$ and $R_{\rm side}$ (see text).}
\vspace{0.25cm}
\begin{tabular}{|l|c||c||c|c|c|}
\hline
$E_{\rm beam}$ \rule[-2mm]{0mm}{6mm} & Cent.~bin & $\tau_{\rm f}$~(fm/c) & $R_{\rm geo}$~(fm)  & $\eta_{\rm f}^2/T_{\rm f}$ (GeV$^{-1}$) & $\langle v_{t} \rangle$/$c$ \\          \hline \hline
          & 1\rule[-2mm]{0mm}{6mm} & 5.95$\pm$0.14 & $6.01_{-0.67}^{+1.05}$ & $3.74_{-1.55}^{+2.85}$ & - \\ \cline{2-6}
          & 2\rule[-2mm]{0mm}{6mm} & 6.34$\pm$0.13 & $10.10_{-2.21}^{+6.90}$ & $15.43_{-7.45}^{+35.29}$ & - \\ \cline{2-6}
          & 3\rule[-2mm]{0mm}{6mm} & 6.49$\pm$0.11 & $9.96_{-1.98}^{+5.22}$ & $12.84_{-5.92}^{+21.98}$ & - \\ \cline{2-6}
40 AGeV   & 4\rule[-2mm]{0mm}{6mm} & 6.63$\pm$0.10 & $12.08_{-2.59}^{+7.65}$ & $19.06_{-8.75}^{+38.23}$ & - \\ \hline 
          & 1\rule[-2mm]{0mm}{6mm} & 6.49$\pm$0.18 & $5.31_{-0.65}^{+1.04}$ & $2.53_{-1.35}^{+2.60}$ & $0.46_{-0.13}^{+0.14}$ \\ \cline{2-6}
          & 2\rule[-2mm]{0mm}{6mm} & 6.98$\pm$0.10 & $5.92_{-0.43}^{+0.57}$ & $3.04_{-0.93}^{+1.33}$ & $0.50_{-0.07}^{+0.07}$ \\ \cline{2-6}
          & 3\rule[-2mm]{0mm}{6mm} & 7.29$\pm$0.09 & $6.23_{-0.41}^{+0.52}$ & $3.03_{-0.83}^{+1.14}$ & $0.50_{-0.06}^{+0.06}$ \\ \cline{2-6}
80 AGeV   & 4\rule[-2mm]{0mm}{6mm} & 7.30$\pm$0.09 & $7.19_{-0.58}^{+0.78}$ & $4.51_{-1.25}^{+1.85}$ & $0.58_{-0.06}^{+0.07}$ \\ \hline 
          & 1\rule[-2mm]{0mm}{6mm} & 7.34$\pm$0.09 & $5.53_{-0.35}^{+0.44}$ & $2.96_{-0.82}^{+1.11}$ & $0.49_{-0.06}^{+0.06}$ \\ \cline{2-6}
          & 2\rule[-2mm]{0mm}{6mm} & 7.46$\pm$0.08 & $5.90_{-0.36}^{+0.44}$ & $3.53_{-0.84}^{+1.13}$ & $0.53_{-0.05}^{+0.06}$ \\ \cline{2-6}
          & 3\rule[-2mm]{0mm}{6mm} & 7.87$\pm$0.06 & $5.86_{-0.22}^{+0.25}$ & $2.45_{-0.47}^{+0.56}$ & $0.46_{-0.04}^{+0.04}$ \\ \cline{2-6}
158 AGeV  & 4\rule[-2mm]{0mm}{6mm} & 8.23$\pm$0.05 & $6.86_{-0.27}^{+0.30}$ & $3.91_{-0.59}^{+0.70}$ & $0.55_{-0.03}^{+0.03}$ \\ \hline 

\end{tabular}
\label{tab:fitresults}
\end{center}
\end{table}

\begin{figure}
   \centering
   \includegraphics[width=15cm]{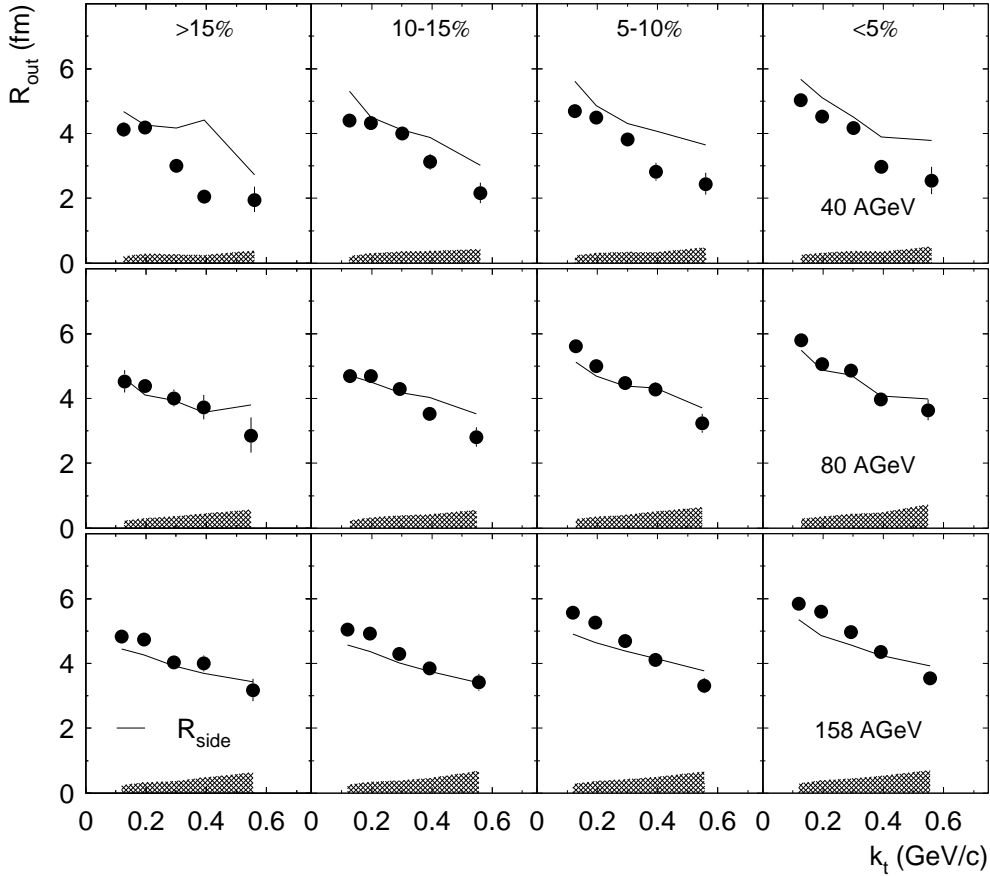}\\
   \caption{The outward radius parameter $R_{\rm out}$ as function
                of $k_{t}$ in different bins of centrality at 40, 80, and
                158 AGeV. The solid line represents the corresponding results for
                $R_{\rm side}$. The shaded regions indicate the systematic error.}
   \label{rout}
\end{figure}

It has been proposed that the existence of a strong first order phase
transistion and an accordingly long lived mixed phase would be 
observable by a large outward radius $R_{\rm out}$ compared to
$R_{\rm side}$, indicating a long duration of pion emission 
$\Delta\tau$~\cite{ber,pra,bergong,berbrown,rischke,rigyu}:
\begin{equation}
  \label{eq:delta_tau}
  \Delta\tau^2 = \frac{1}{\beta^2_t}(R^2_{\rm out}-R^2_{\rm side}),
  \label{deltatau}
\end{equation}
where $\beta_t$$\approx$$k_t/m_t$ is the average transverse velocity of the 
pion pair in the LCMS ($\beta_l$$\approx$0).
The dependence of the $R_{\rm out}$ parameter on $k_t$ 
is shown in Fig.~\ref{rout}. 
%The shaded band indicates the size 
%of the correction which was applied to account for the finite
%momentum resolution.
For comparison, $R_{\rm side}$ is 
also shown by the solid lines. 
We observe a smooth increase of $R_{\rm out}$ with centrality and
beam energy. At 158 AGeV, the ratio $R_{\rm out}$/$R_{\rm side}$
is larger than unity at small $k_t$. While $R_{\rm out}$/$R_{\rm side}$$<$1 was observed
at large $k_t$ recently at RHIC, we see at the highest beam energy
a ratio consistent with one, within the systematic uncertainties
which are mainly imposed by the large corrections to $R_{\rm out}$. 

The data do not support the scenario of a long-lived source at SPS: All observed
radii $R_{\rm out}$ are similar to $R_{\rm side}$ at all beam
energies and centralities.
At the lowest beam energy $R_{\rm side}$ 
is even slightly larger than $R_{\rm out}$.
We note that $R_{\rm out}$$<$$R_{\rm side}$ was suggested
as an indication for sources with surface dominated 
emission~\cite{heisel}, such as emission from an expanding shell.
At 158 AGeV, the data are consistent with a short but finite
emission duration of about 2-3~fm/c, in agreement with results reported by NA49
previously~\cite{na49hbt}. 

\begin{figure}
   \centering
   \includegraphics[width=15cm]{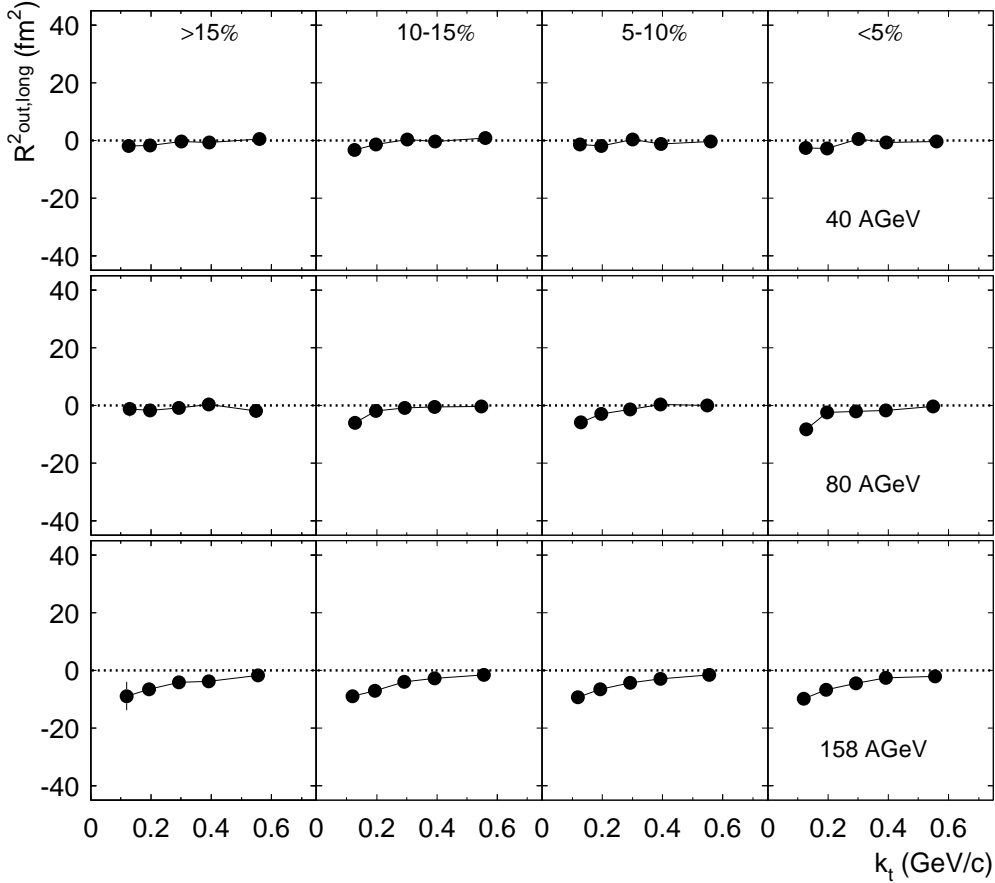}\\
   \caption{The cross terms $R^{2}_{\rm out,long}$ as function
                of $k_{t}$ in different bins of centrality at 40, 80, and
                158 AGeV.}
   \label{routlong}
\end{figure}

The existence of a non-vanishing cross term $R_{\rm out,long}^{2}$ 
would indicate space-time correlations of the pion emission
points~\cite{heinzrol}. Such correlations occur if the longitudinal
flow velocity of the source element with highest emissivity 
does not coincide with the average longitudinal velocity of
the pions. For symmetry reasons, $R_{\rm out,long}^{2}$
is expected to vanish at midrapidity and in sources with
longitudinal boost-invariant expansion. Our results 
for the cross term $R_{\rm out,long}^{2}$ are shown
in Fig.~\ref{routlong}. While $R_{\rm out,long}^{2}$
is consistent with zero at 40 AGeV, we observe a smooth
increase of its absolute value with beam energy at small $k_t$. 
The small but non-zero results at the higher beam energies
may be caused by a slight shift of the rapidity acceptance
of the spectrometer away from midrapidity as the beam energy increases.

\begin{figure}
   \centering
   \includegraphics[width=15cm]{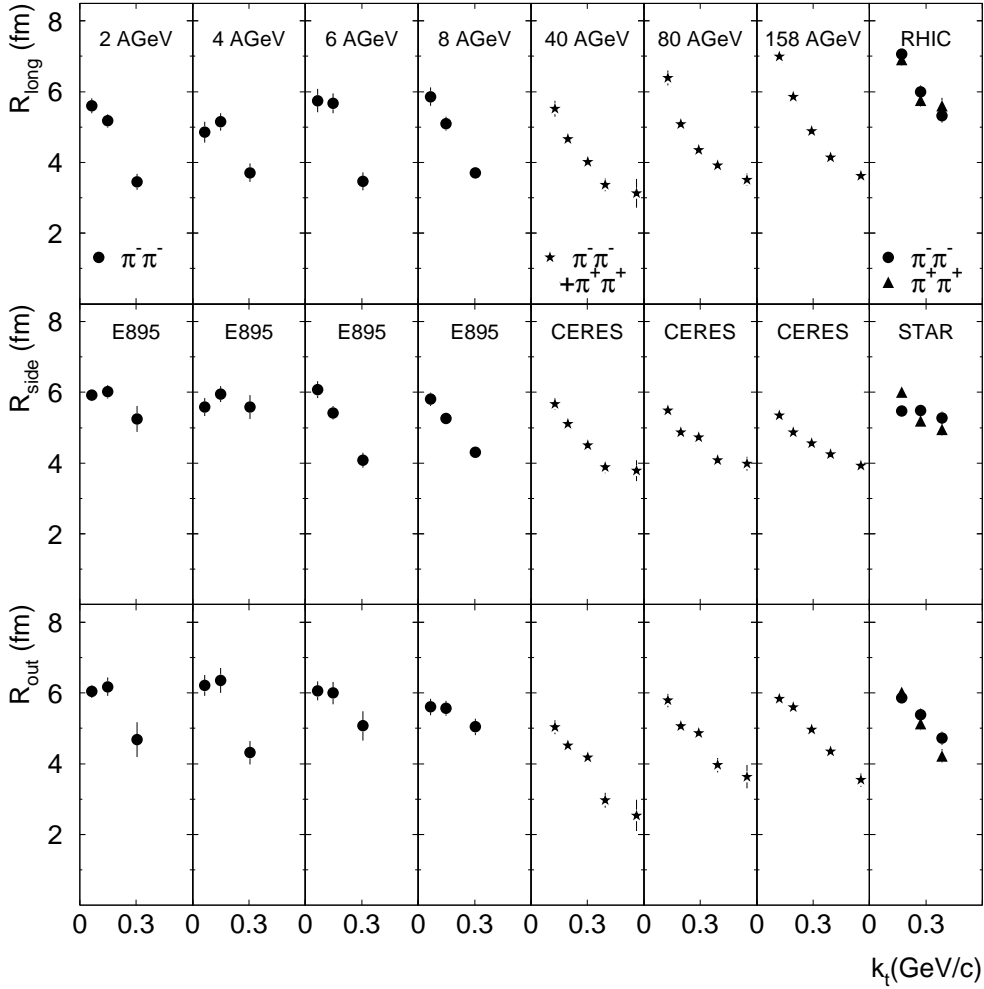}\\
   \caption{Compilation of HBT radius parameters near midrapidity 
        in central Pb(Au)+Pb(Au) collisions
        at AGS~\cite{e895hbt}, SPS (present data), and RHIC~\cite{starhbt} energies.}
   \label{edep}
\end{figure}
Pion interferometry data published by experiments
at AGS and RHIC together with the data presented here
can be combined to perform a systematic study of the source parameters
over a wide range of beam energies.
In Fig.~\ref{edep} are shown the $k_{t}$-dependences of $R_{\rm long}$, $R_{\rm side}$,
and $R_{\rm out}$ in central Pb(Au)+Pb(Au) collisions near midrapidity~\cite{e895hbt,starhbt}.
No dramatic variation of the source parameters 
can be observed.
However, a closer inspection reveals interesting features. The parameter $R_{\rm long}$
is approximately constant from AGS to the lower SPS energies, but starts to
increase significantly within the SPS regime and towards RHIC, indicating a smooth
increase of the lifetime.
$R_{\rm side}$ is gradually decreasing
at small $k_{t}$ up to top SPS energy, connected with a continuous flattening
of the $k_{t}$-dependence. At RHIC, $R_{\rm side}$ is again larger than at the SPS
while the shape is not yet well measured. In fact, the $\pi^{-}\pi^{-}$ sample
indicates further flattening.
Naively, the flattening would indicate a decrease of the radial flow velocity
as function of beam energy. This is in contradiction to the present
interpretation of single particle transverse mass spectra,
which indicate an increase of radial flow with beam energy~\cite{xuqm}.
The parameter $R_{\rm out}$ shows a rather weak energy dependence
and a slight minimum around the lowest SPS energy, where we find 
$R_{\rm out}$/$R_{\rm side}$$<$1.

\begin{figure}
   \centering
   \includegraphics[width=14cm]{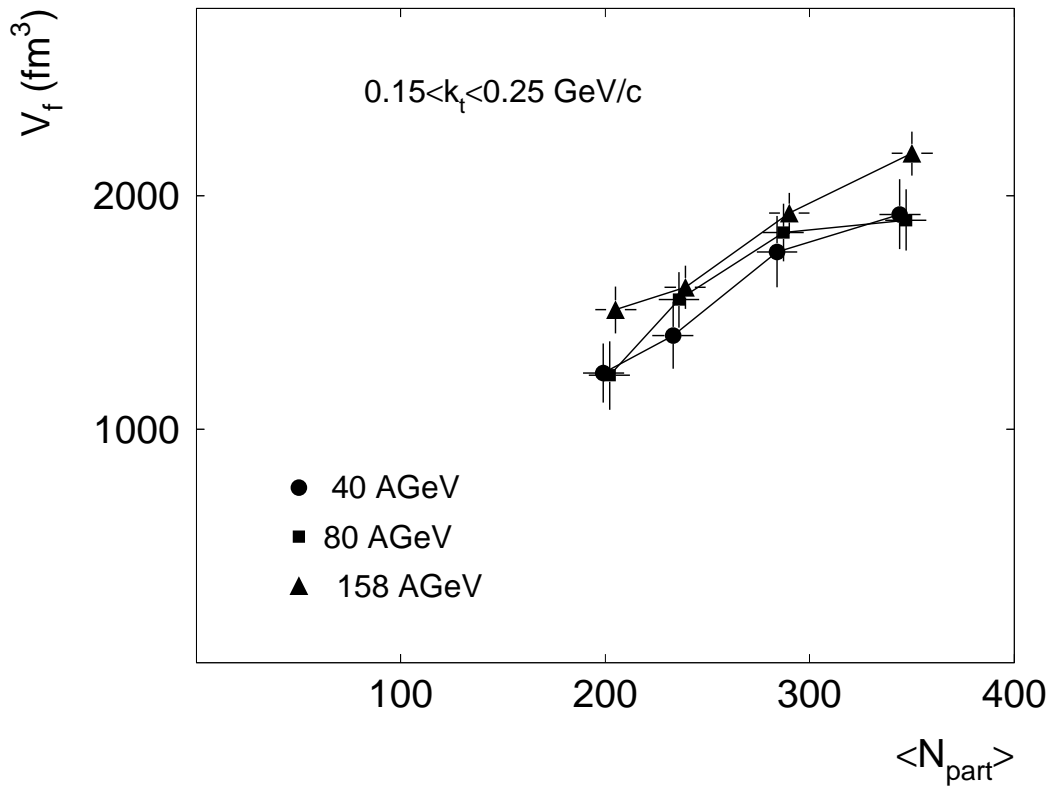}\\
   \caption{The freeze-out volume $V_f$ calculated at
        0.15$<$$k_{t}$$<$0.25~GeV/c (see text) as function of the
        centrality expressed here in terms of the 
        number of participants at 40, 80, and 158 AGeV.}
   \label{vol_cent}
\end{figure}

An investigation of the freeze-out conditions can
be performed by relating the measured source parameters to an effective
freeze-out volume: 
\begin{equation}
V_f=(2\pi)^{\frac{3}{2}} R_{\rm long}R_{\rm side}^{2},
\end{equation}
computed for Gaussian density distributions in all three
spatial dimensions.
In the presence of collective expansion, $V_f$ does not comprise the 
total volume of the pion source at freeze-out 
but rather reflects a volume of homogeneity. 
If thermal freeze-out would occur at constant density~\cite{pom},
a linear scaling of $V_f$ with the charged particle
multiplicity would be expected.
Fig.~\ref{vol_cent} shows
$V_f$, computed at 0.15$<$$k_t$$<$0.25~GeV/c, 
as function of the number of participants at 40, 80, and 158 AGeV.
An approximately linear scaling with $N_{\rm part}$ is indeed observed at all 
three energies. 
Since the number of
charged particles was found to scale close to linear with $N_{\rm part}$ at
SPS~\cite{wa98mult,na45mult,wa97mult}, this is
consistent with the assumption of a constant freeze-out density at a given beam energy.
However, the observed beam energy dependence is at first glance surprising: 
the increase of pion multiplicity at midrapidity by about 70\%
from 40~AGeV to 158~AGeV~\cite{na49spectra} is not reflected in a corresponding 
increase of $V_f$.
Obviously, the freeze-out
volume scales with multiplicity as long as multiplicity is controlled via
centrality, but it does not scale accordingly as multiplicity changes with
beam energy. 
We note that, according to experimental results, the ratios of particle 
species change very little with centrality for not too peripheral collisions,
whereas they change significantly with beam energy.
In a separate publication~\cite{na45letter} we show that thermal pion freeze-out
in general does not occur at constant density $\rho_f$$=$$N$/$V_f$, if relative
abundancies of hadrons change. 
This is due to the different cross sections of pions with e.g.~pions and protons,
which explains that freeze-out occurs 
rather at constant mean free path $\lambda_f$=$(\rho_f\cdot\sigma_{\rm eff})^{-1}$, 
where $\sigma_{\rm eff}$ is the effective cross-section
of the system with pions, averaged over all particle species in the final
state: $\sigma_{\rm eff}$=1/$N\cdot\sum_{i}{N_{i}\sigma_{\pi i}}$.

\section{SUMMARY AND CONCLUSIONS}
We have presented a systematic study of two-pion interferometry
data at SPS energies. We observe a smooth centrality and beam energy dependence
of all HBT parameters; no indications for drastic changes of the space-time
evolution have been found. The data at all centralities and beam energies
under investigation suggest the existence of a pion source 
which rapidly expands in longitudinal and transverse direction. 
Assuming longitudinal boost-invariance, a duration of expansion
of about 6-8~fm/c can be determined from the pair transverse 
momentum dependence of the longitudinal radius parameter $R_{\rm long}$,
smoothly increasing with centrality and beam energy. The expansion
is followed by a rather sudden freeze-out with a short duration of pion
emission. At the time of freeze-out, the transverse r.m.s.~ radius 
of the system has
increased by about a factor two compared to the dimension of the initial overlap
region of the two colliding nuclei. 
This implies that the transverse area of the reaction has quadrupled.
The transverse momentum dependence
of the parameter $R_{\rm side}$ indicates a high degree of collectivity 
and a radial flow velocity of about $\langle v_{t} \rangle$$\approx$0.55$c$
in the most central collisions. 

Studying the beam energy dependence of $R_{\rm long}$,
there is a monotonic increase 
from top AGS energies to RHIC. 
Over the same range of beam energies there is a subtle but systematic
flattening of the $k_t$-dependence of $R_{\rm side}$.
More studies are needed to understand the origin of this effect.
The centrality dependence of the effective freeze-out volume $V_{f}$
is consistent with the assumption of pion freeze-out at constant
particle density. However, this simple picture breaks down if
different beam energies are compared. The key to understand
this effect is in the consideration of the cross sections
of different particle species with pions, as the relative abundances
change with beam energy.

\section*{Acknowledgements}
The CERES collaboration acknowledges the good performance of the CERN
PS and SPS accelerators as well as the support from the EST division. 
We are grateful for excellent support for the central data recording
from the IT division. 
For guidance in the design of the TPC we thank V.~Eckardt, W.~Klempt,
E.~Rosso, and B.~Goret as well as G.~Augustinski, A.~Przybyla, and R.~Ziegler 
for technical help.
The new TPC readout in 2000 was only made possible through the
strong support of
F.~Formenti, M.~Jost, and G.~Thomas.
We wish to thank P.~Filip, U.~Heinz, P.~Seyboth, and B.~Tom\'{a}\v{s}ik
for valuable discussions. 
This work was supported by the German BMBF, the U.S.~DoE, the 
Israeli Science Foundation, and the MINERVA Foundation.

% The Appendices part is started with the command \appendix;
% appendix sections are then done as normal sections
% \appendix

% \section{}
% \label{}

\end{document}